\documentclass[twocolumn,english,keywords]{revtex4}
\usepackage[T1]{fontenc}
\usepackage[latin1]{inputenc}
\usepackage{amsmath}
\usepackage{amssymb}

\makeatletter
\usepackage[T1]{fontenc}
\usepackage[latin1]{inputenc}
\makeatletter
\usepackage[T1]{fontenc}
\usepackage[latin1]{inputenc}
\makeatletter

\usepackage[T1]{fontenc}
\usepackage[latin1]{inputenc}
\makeatletter

\usepackage[T1]{fontenc}
\usepackage[latin1]{inputenc}
\usepackage{amssymb}

\makeatletter

\usepackage{graphicx}
\usepackage{amsfonts}
\usepackage{amssymb}
\usepackage{babel}
\addto\extrasspanish{\bbl@deactivate{~}}
\usepackage{babel}
\makeatother

\usepackage{babel}
\makeatother

\makeatother

\usepackage{babel}
\makeatother

\usepackage{babel}
\makeatother
\begin{document}

\title{The Inverse Variational Problem and Logistic Self-Regulated Systems }

\author{\'{A}. G. Mu\~{n}oz S.$^{1,2}$%
\footnote{To whom the correspondence should be adressed. E-mail: agmunoz@mail.luz.ve%
}, D. Sierra Porta$^{1,2}$, T. Soldovieri$^{1}$}

\address{$^{1}$Grupo de Investigaciones de Física Teórica (GIFT) Departamento
de Física. Facultad de Ciencias. La Universidad del Zulia. Maracaibo
4004, Venezuela.}

\address{$^{2}$Postgrado en Física Fundamental, Centro de Astrofísica Teórica
(CAT), Facultad de Ciencias, Universidad de Los Andes. Mérida, 5101,
Venezuela.}

\begin{abstract}
There exists in nature many examples of systems presenting self-limiting
behaviour: population dynamics, structure engineering, Townsend's
electron breakdown, nuclear decay in radioactive equilibrium, histeresis
process, meteorological models, etcetera. In this work we call attention
to the advantages the use of a variational formulation should provide
to the study of self-regulated systems, such as a unified description
of the related phenomena, further comprehension of the internal structure
and symmetries of the related equations, and the determination of
the equilibria points via the energy function. We study the case of
logistic systems, obtaining explicitly several s-equivalent Lagrangeans
corresponding to the Verhulst's equation. Some dynamical properties
are discussed in the light of this approach. Expressions for the mean
energy function are also obtained. 

Keywords: logistic equation, Verhulst's Lagrangean, self-regulated
systems.

PACS: 02.30.Zz, 45.20.Jj, 87.23.Cc
\end{abstract}
\maketitle

\section{Introduction}

The Lagrangean formalism is a very powerful tool in physics, presenting
in a single formula all the dynamical information of a system. Nevertheless,
it is widely accepted that a variational principle cannot be constructed
for an arbitrary differential equation \cite{Van}: there exists a
strict mathematical theorem that shows its existence for a given situation,
and whose application reduces the number of equations in physics that
have a Lagrangean-Hamiltonian formulation \cite{Veinberg,Tonti}.
Due to its importance, efforts have been directed to make several
methods to circumvent this condition and to construct modified variational
principles \cite{Van,Ichi,VanMu,Nyi}.

However, there is an interesting different approach, known as the
inverse problem of the variational calculus, or IPVC (see, for example,
\cite{Hojman-Urrut,HojmaShep}), that has its historical origin in
the times of Helmholtz \cite{Helm}. It consists in studying the existence
and uniqueness (or multiplicity) of Lagrangeans for systems of differential
equations, meaning finding the Lagrangean, if it exists, from the
equations of motion, instead of the traditional approach. This proceedure
has become quite useful and one important result is that of Hojman
et al. \cite{Hojman}, who have proved that it is possible to construct
the Lagrangean for any regular mechanical system as linear combination
of their own equations of motion. This particular construction, for
example, is much wider than the traditional definition $L=T-V$, which
is only true when the {}``forces'' involved are derivable from position-dependent
potentials (or very few cases of velocity-dependent potentials), therefore
it may be used for general non-conservative systems.

Whereas the equations of the models employed to describe auto-regulated
phenomena can be understood as equations of motion in the variational
sense, the latter approach shows itself to be one of the most adequate
to obtain the Lagrangean formulation of the problem. Its application
to the study of self-limiting processes may provide additional understanding
into the internal structure of these phenomena, also enables the use
of a well known mathematical machinery to find conserved quantities,
equilibria and stability cases, and other dynamical properties. 

In particular, the method proposed by Hojman et al. provides the desired
Lagrangean both elegantly and without many complications, relating
directly the constants of motion with the problem. Bearing this in
mind, in the following pages we treat the problem of self-regulated
systems described by the Verhulst's Logistic Equation (VLE, from now
on) \cite{verhulst,murray} by means of the Hojman proceedure, showing
how a variational formulation can be easily obtained once we know
the equation of motion and some additional information about the system.
We have chosen the VLE because it had been well studied, thus we can
easily compare the results here obtained to evaluate the approach,
and also becasuse the model is simple enough as to give an idea, without
involving too much time, of the way a self-regulatory system works.

Historically this equation, firstly proposed by P.F. Verhulst \cite{verhulst}
in 1838 as a continuous non-linear population growth model with a
self-limiting density dependent mechanism, was suggested as a direct
generalization of the Euler and Malthus \cite{malthus} studies about
population dynamics.

If $n(t)$ is the population at a time $t$, $A^{-1}$ the representative
time scale of response of the model to any change in the population
and $B$ define the carrying capacity of the enviroment (see below),
then the logistic equation can be written as \begin{equation}
\frac{dn}{dt}=An\left(1-\frac{n}{B}\right)\label{verh1}\end{equation}
 The solution of equation (\ref{verh1}), for $n(0)=n_{0}$ is: \begin{equation}
n(t)=\frac{n_{0}Be^{At}}{B+n_{0}\left(e^{At}-1\right)}\label{ver}\end{equation}
 and we can inmediatly show that $\lim_{t\rightarrow\infty}n(t)=B$
(for all $n_{0}$), meaning, of course, that the carrying capacity
corresponds to the the maximum size of the stable steady state population$.$
In the Figure 1 the qualitative behavior of (\ref{ver}) is presented.
For $n_{0}<B$ the profile is the characteristic sigmoid of the model,
and for $n_{0}>B$ the behavior is similar to an exponentially decay
function.

Indeed, Verhulst knew \cite{darcy} that there were {}``biological
obstacles'' in the natural population growth, but to write a general
mathematical expression for them was hardly possible, because each
obstacle could affect in a different way the growth itself. Instead,
he proposed a general {}``retardatrice'' density dependent ($\propto n^{2}$)
function, a kind of compensating effect of overcrowding. However,
it is important to remark that, even a very simple model, it is really
useful as it matches a wide variety of different contexts, not only
in population dynamics \cite{murray}, but also in biological growth
(see for example, \cite{darcy}), radioactive decay series \cite{Aonso},
and even in some reaction-diffusion models \cite{Nelson}. Effectively,
all these cases (and more) can be adequately described by means of
a characteristic saturation curve like (\ref{ver}) and in these cases
an interpretation of the phenomenona by means of change in the population
of some element (individuals, cells, isotopes, ...) can be made.

The self-regulated systems, not only those described by the VLE, are
present everywhere: certain aspects in stability of structures (firstly
pointed out by Euler \cite{Euler}), Townsend electron breakdown \cite{Town},
histeresis and magnetization processes \cite{Joos}, the Amdhal law
for scalability of computer programs and a long etcetera. As far as
these phenomena are studied sometimes by quite different disciplines
of science, it is interesting to explore if it is possible to provide
a unified description for them in terms of families of Lagrangeans
(or Hamiltonians), that also would help to classify the systems by
their dynamical properties. 

The distribution of this work is the following: Next section contains
an abstract of the Hojman et al. method, which we use in Section 3
to construct the Verhulst's Lagrangian and Hamiltonian. Section 4
is devoted to the study of the energies related to the system, its
equilibrium points and its constant of motion. Finally the concluding
remarks are presented in Section 5.

\section{The Hojman et al. Method}

For the purposes of this paper it is useful to work with the usually
denominated second order approach \cite{Hojman}. In this context,
the equation of motion for a mechanical system arises from a set of
$m$ differential equations: \begin{equation}
G^{i}\equiv\overset{..}q^{i}-F^{i}\left(q^{j},\overset{.}q^{j},t\right)=0,\,\,\,\,\,\,\,\,\,\,\,\, i,j=1,...,m\label{G}\end{equation}
 where the $q^{i}$ are the generalized coordinates and the point
means total temporal derivative.

In the IPVC the Lagrangean $L\left(\overset{.}q^{j},q^{j},t\right)$
is constructed such that relations (\ref{G}) can be effectively deduced
via the Euler-Lagrange equations. The existence of such a Lagrangean
is studied by means of the nowdays called Helmholtz conditions \cite{pardo}\begin{equation}
\frac{\partial G_{i}}{\partial\overset{..}q^{j}}=\frac{\partial G_{j}}{\partial\overset{..}q^{i}}\end{equation}
\begin{equation}
\frac{\partial G_{i}}{\partial\overset{.}q^{j}}+\frac{\partial G_{j}}{\partial\overset{.}q^{i}}=\frac{d}{dt}\left(\frac{\partial G_{i}}{\partial\overset{..}q^{j}}+\frac{\partial G_{j}}{\partial\overset{..}q^{i}}\right)\end{equation}
\begin{equation}
\frac{\partial G_{i}}{\partial q^{j}}-\frac{\partial G_{j}}{\partial q^{i}}=\frac{1}{2}\frac{d}{dt}\left(\frac{\partial G_{i}}{\partial\overset{.}q^{j}}-\frac{\partial G_{j}}{\partial\overset{.}q^{i}}\right)\end{equation}

Nevertheless, these conditions do not give any warranty about uniqueness.
Two Lagrangeans are said to be solution-equivalents (or s-equivalents)
if they just differ by a global multiplicative constant, $\eta$,
and a total time derivative of some gauge $\Lambda\left(\overset{.}q^{j},q^{j},t\right)$:\begin{equation}
\eta L=\overset{\thicksim}{L}+\frac{d\Lambda}{dt}\label{Lhoj}\end{equation}

The different systems of equations they provide, however, have exactly
the same equations of motion.

We can proceed now to describe briefly the method. It enables us to
write $\overset{\thicksim}{L}$ as a linear combination of the known
equations of motion; then for $i,j=1...m$ (the degrees of freedom),\begin{equation}
\overset{\thicksim}{L}=\mu_{i}\left[\overset{..}q^{i}-F^{i}\left(q^{j},\overset{.}q^{j},t\right)\right]\label{lbarra1}\end{equation}
 where\begin{eqnarray}
\mu_{i}\left(q^{j},\overset{.}q^{j},t\right) & \equiv & D_{1}\frac{\partial D_{2}}{\partial\overset{.}q^{i}}+...+D_{2m-1}\frac{\partial D_{2m}}{\partial\overset{.}q^{i}}\label{mul}\\
 & = & -\frac{\partial\Lambda}{\partial\overset{.}q^{i}}\end{eqnarray}
 and the $D_{2m}$ are constants of motion of the mechanical system
and the $D_{2m-1}$ are arbitrary functions whose arguments are constants
of motion. One possible form for the $D_{2m-1}$ functions, given
the $D_{2m}$ conserved quantities, is presented in reference \cite{Hojman-Urrut}.

When the conserved quantities are unknown then the problem is reduced
to find $\mu_{i}$ such that the following system is satisfied:\begin{equation}
\left\{ \begin{array}{c}
\frac{\partial\mu_{i}}{\partial\overset{.}q^{j}}=\frac{\partial\mu_{j}}{\partial\overset{.}q^{i}}\\
\frac{\overset{\_}{d}}{dt}\left(\frac{\overset{\_}{d}}{dt}\mu_{i}+\mu_{j}\frac{\partial F^{j}}{\partial\overset{.}q^{i}}\right)-\mu_{j}\frac{\partial F^{j}}{\partial\overset{.}q^{i}}=0\end{array}\right.\label{SisHoj}\end{equation}
 and\begin{equation}
\det\left[\frac{\partial}{\partial\overset{.}q^{j}}\left(\frac{\overset{\_}{d}}{dt}\mu_{i}+\mu_{k}\frac{\partial F^{k}}{\partial\overset{.}q^{j}}\right)+\frac{\partial\mu_{i}}{\partial q^{j}}\right]\neq0\end{equation}
 where the on-shell derivative $\frac{\overset{\_}{d}}{dt}$ is defined
as\begin{equation}
\frac{\overset{\_}{d}}{dt}\equiv F^{i}\frac{\partial}{\partial\overset{.}q^{i}}+\overset{.}q^{i}\frac{\partial}{\partial q^{i}}+\frac{\partial}{\partial t}\end{equation}

For further details the reader is exhorted to review \cite{Hojman-Urrut,Hojman}
and the references therein. In the following section we make use of
this proceedure to obtain the Lagrangean corresponding to the VLE.

\section{Constructing the Verhulst's Lagrangean}

For convenience, we shall write (\ref{verh1}) as \ref{vermod}\begin{equation}
\overset{.}{q}\equiv\frac{dq}{dt}=kq\left(B-q\right)\label{vermod}\end{equation}
 with $k\equiv A/B$, and we shall consider the one-dimensional problem.

In our case, equation (\ref{G}) is\begin{equation}
\overset{..}{q}-k\overset{.}{q}(B-2q)=0\end{equation}
 where $\overset{.}{q}$ is given by (\ref{vermod}). Explicitly in
terms of the generalized coordinate (on shell), we have\begin{equation}
\overset{..}{q}-k^{2}q(B-q)(B-2q)=0\label{ecumov}\end{equation}
 together with the initial conditions\begin{equation}
\overset{.}{q}(0)=kq_{0}\left(B-q_{0}\right)\end{equation}

\begin{equation}
q(0)=q_{0}\end{equation}

Note at this point that, indeed, the Verhulst Lagrangean exist, for
equation (\ref{ecumov}) satisfies trivially the Helmoltz conditions.

The second order Hojman et al. method provides then\begin{equation}
\overset{\thicksim}{L_{V}}=\mu\left[\overset{..}{q}-k^{2}q(B-q)(B-2q)\right]\label{Lvbarn}\end{equation}

We just need to determine the factor $\mu$, which in this case can
be written as\begin{equation}
\mu=C_{1}\frac{\partial C_{2}}{\partial\overset{.}{q}}\label{muc2}\end{equation}

There are several ways to proceed now. For example, observe that from
the initial conditions and (\ref{ecumov}) we can obtain the only
constant of motion, $C_{2}$, the system possesses (of course, for
$m$ degrees of freedom there are $(2m-1)$ functionally independent
constans of motion \cite{Landau}); the arbitrary function $C_{1}$
has then a prescribed form \cite{Hojman-Urrut}, and $\mu$ is obtained
by means of (\ref{muc2}). Furthermore, it is possible to solve (\ref{SisHoj})
in the one-dimensional case for $\mu$, which yields a general solution
in terms of Bessel functions. Using the initial conditions of the
problem we find the particular solution. It is also possible to introduce
the expression for the Lagrangean in the Euler-Lagrange equation and
compare with the equation of motion to obtain $\mu$%
\footnote{Even when not indicated by Hojman and colaborators, this idea may
have some advantages. An adequate choice of the gauge and the use
of equation (\ref{Lvbarn}) and (\ref{Lhoj}) can simplify sometimes
the calculations and the final Lagrangean.%
}. We shall ilustrate here the first way.

It is not difficult to show that \begin{equation}
C_{2}\equiv\frac{1}{2}\overset{.}q^{2}-\frac{1}{2}k^{2}q^{2}\left(B-q\right)^{2}=0\end{equation}
and a possible choice for $C_{1}$ is (see reference \cite{Hojman-Urrut})\begin{equation}
C_{1}=cC_{2}^{2}\label{C1}\end{equation}
where $c$ is an arbitrary constant that multiplies the equation of
motion. The Lagrangean reads then\begin{equation}
\overset{\thicksim}{L_{V}}=cC_{2}^{2}\overset{.}q\left[\overset{..}{q}-k^{2}q(B-q)(B-2q)\right]\label{LbarVerf}\end{equation}

When inserted into the Euler-Lagrange equation, (\ref{LbarVerf})
provides the corresponding equation of motion (\ref{ecumov}) plus
terms that are zero in virtue of the constant of motion. However,
it has been shown that for the one-dimensional problem there exists
an infinite number of Lagrangeans \cite{HojmaShep,Hojman}. Thus,
we can write a more simpler s-equivalent one by means of the total
time derivative of a certain gauge. Given the constant of motion,
let choose\begin{equation}
\frac{d\Lambda}{dt}=\overset{.}q^{2}-\left(C_{1}\frac{dC_{2}}{dt}+C_{2}\right)\label{difcalib}\end{equation}

Without loss of generality we set $c=\eta=1$, and together with (\ref{difcalib}),
equations (\ref{Lhoj}) and (\ref{LbarVerf}) provide\begin{equation}
L_{V}=\frac{1}{2}\overset{.}q^{2}+\frac{1}{2}k^{2}q^{2}\left(B-q\right)^{2}\label{lvfin}\end{equation}
 and the Hamiltonian can be written as \begin{equation}
H_{V}=\frac{1}{2}P^{2}-\frac{1}{2}k^{2}q^{2}\left(B-q\right)^{2}\label{hv}\end{equation}
 where the generalized momentum is \begin{equation}
P\equiv\frac{\partial L}{\partial\overset{.}{q}}=\overset{.}{q}\end{equation}

\section{Conservation, Energies and Equilibrium Conditions }

Now we proceed to study some of the applications the Lagrangean formalism
offers to the Verhulst system.

The first term in (\ref{lvfin}) and (\ref{hv}) is identified as
the usual traslational {}``kinetic'' energy: quadratic and homogeneous
in the first temporal derivative of the generalized coordinate. The
second term on the right side of equation (\ref{hv}) is explicitily
time-independent, meaning that $H_{V}$ coincides with the energy
function and it is then possible to write the potential energy of
the system as \begin{equation}
V=-\frac{1}{2}k^{2}q^{2}\left(B-q\right)^{2}\label{poten}\end{equation}

The behavior of this potential is qualitatively presented in Figure
2 for fixed carrying capacity. In this potential, the smaller the
time scale ($1/A$), the more closed the curve and the larger the
depression. Likewise, the greater the carrying capacity, $B$, the
more pronounced the depression and the more open the curve. Equation
(\ref{poten}) can be interpreted as an impulsor-retardatrice potential,
depending only on the generalized coordinate. In fact, observe that
when $q_{0}\leq B$, for the interval $\left(0,B/2\right)$ the acceleration
associated impulses the movement, being conversely for the interval
$\left(B/2,B\right)$; the point $B/2$ corresponds to a local minimum,
where the acceleration instantaneously annuls itself. On the other
hand, when $q_{0}>B$, i.e. the interval $\left(B,\infty\right)$,
the character of the acceleration is always impulsive, even being
$\overset{.}{q}<0$. All this is consistent with the profiles sketched
in Figure 1 and with the standard knowledge about equation (\ref{vermod}).

Despite the fact the VLE describes multiple phenomena which one should
think presents dissipation of energy (for example as in population
dynamics, growth of living beings or meteorological models), it is
easy to show that the Verhulst system is conservative (population
systems with a first integral has been studied, see \cite{murray}
and references therein); thus, the energy function $H_{V}$ equals
the total energy $E$. Note that in virtue of (\ref{vermod}) equation
(\ref{hv}) is identically zero. Taking all these considerations into
account, it is clear in Figure 2 that the kinetic energy tends to
annul itself as $q$ approachs $B$. As a consequence, the system
takes an infinite time to reach the steady state, and $q_{0}<B$ implies
always $q<B$ (similarly for $q_{0}>B$).

Another interesting issue is that there are not oscillatory behaviours,
despite the similarity with harmonic or the anharmonic potentials.
Observe that as far as we are just interested in feassible results,
we must take only the positive values of $q(t)$. Thus the character
of the force (see equation (\ref{ecumov})) is not restorative, but
the commented before for each interval.

One important application is that of finding the stability cases.
The usual derivative criteria allows us to find the equilibrium points,
for which \begin{equation}
\frac{\partial E}{\partial q}=0\label{condequ}\end{equation}
 The stability is then ruled by the condition \begin{equation}
\frac{\partial^{2}E}{\partial q^{2}}>0\end{equation}

The equilibrium solutions are $q_{e}=\left\{ 0,\frac{B}{2},B\right\} $.
The first one corresponds to a point of instability, the second to
minimal potential energy and the last one is, as expected, the state
of stationary equilibrium. As before, this analysis is in accordance
with the known behaviour of the VLE (see Figure 1). It is important
to remark here that usually the study of its equilibrium points is
treated by means of a Taylor expansion of (\ref{vermod}) about $q=0$
and $q=B$ (see for example \cite{murray}). The variational formalism
provides thus both a more elegant and complete treatment.

On the other hand, as far as the VLE can model various self-regulated
phenomena, it should be useful for the description of the related
statistical systems to write the total mean energy. We shall use the
virial theorem for this porpouse. In our case, \begin{eqnarray}
\left\langle T\right\rangle  & = & -\frac{1}{2}\left\langle \frac{\partial V}{\partial q}q\right\rangle \nonumber \\
 & = & \frac{1}{2}k^{2}\left\langle q^{2}\left(B-q\right)\left(B-2q\right)\right\rangle \label{tmed}\end{eqnarray}

Then, the total mean energy is written as \begin{eqnarray}
\left\langle E\right\rangle \equiv & \left\langle T+V\right\rangle  & =-\frac{1}{2}k^{2}\left\langle q^{3}(B-q)\right\rangle \nonumber \\
 &  & =-\frac{k}{6}\left\langle \frac{d}{dt}\left(q^{3}\right)\right\rangle \label{etotal}\end{eqnarray}
or, explicitily, evaluating the integral between $t=0$ and $t\rightarrow\infty$,\begin{eqnarray}
\left\langle E\right\rangle _{\infty} & = & \lim_{T\rightarrow\infty}\left\{ -\frac{k}{6T}\left[q(T)^{3}-q_{0}^{3}\right]\right\} \nonumber \\
 & = & 0\label{etot0}\end{eqnarray}
 Thus the total mean energy of the systems described by VLE is always
null. But if as superior temporal bound we choose the characteristic
time of the system, then\begin{equation}
\left\langle E\right\rangle _{1/A}=-\frac{A^{2}}{6B}\left[q\left(\frac{1}{A}\right)^{3}-q_{0}^{3}\right]\end{equation}

Returning to the general case, observe that under certain circumstances
equation (\ref{etotal}) make it possible to stablish a relationship
between the mean-squared velocity and the mean-linear velocity. Indeed,
solving (\ref{vermod}) for $q$ we find two solutions:\begin{equation}
q_{\pm}(t)=\frac{kB\pm\sqrt{B^{2}k^{2}-4k\overset{.}q}}{2k}\end{equation}
 Using $q_{+}(t)$, equation (\ref{etotal}) can be restated as\begin{equation}
\left\langle E\right\rangle =\left\langle \frac{\overset{.}q^{2}}{2}-\frac{B}{4}\overset{.}q\left[Bk+\sqrt{B^{2}k^{2}-4k\overset{.}q}\right]\right\rangle \end{equation}
 Then\begin{equation}
\left\langle \overset{.}q^{2}\right\rangle =\frac{B}{2}\left\langle \overset{.}q\left[Bk+\sqrt{B^{2}k^{2}-4k\overset{.}q}\right]\right\rangle +2\left\langle E\right\rangle \label{velcmed}\end{equation}
 Note that if $\left|\frac{4\overset{.}q}{B²k}\right|\ll1$ the latter
equation simplifies to\begin{equation}
\left\langle \overset{.}q^{2}\right\rangle =\frac{AB}{2}\left\langle \overset{.}q\right\rangle +2\left\langle E\right\rangle \end{equation}

The use of the other solution, $q_{-}(t)$, provides formally\begin{equation}
\left\langle E\right\rangle =0\end{equation}
and then equation (\ref{velcmed}) is no longer true.

\section{Concluding Remarks}

We suggest in this work the use of the IPVC to obtain a variational
formulation for self-limiting systems. This approach provides at least
a Lagrangean, if it exists, departing from the equation of motion
and regardless the system being conservative (when the definition
$L=T-V$ holds) or not. There are several advantages, namely, it is
possible 

(a) to have in just one relation all the dynamical information of
the system; 

(b) to use a well know mathematical machinery to obtain from the Lagrangean
or Hamiltonian the conserved quantities (and symmetries), equilibria
points and trajectories, etc.;

(c) to group self-regulated phenomena having the same variational
formulation, even when describing quite different scenarios, classifying
the systems by the form of the Lagrangean or by its dynamical properties; 

(d) to solve, when possible, the equation of motion. For some systems
(as those described by Lotka-Volterra models) the Hamilton-Jacobi
formalism, for example, should be useful to integrate the equation
of motion;

(e) to help to understand how the models are related. Some models
are, indeed, particular cases of more general equations;

(f) to test if a certain model is acceptable in terms of the Least-Action
Principle. 

On the other hand, making use of the Hojman et al. method, we have
found the Lagrangean (\ref{lvfin}) and Hamiltonian (\ref{hv}) corresponding
to the (1+1)-dimensional logistic equation as an example of the way
a variational formalism can be obtained for general self-regulated
systems. This may be regarded as a proof that the VLE corresponds
to an extremal trajectory, thus it is physically acceptable (it complies
with the Hamilton's Least-Action Principle) despite the fact Verhulst
introduced it heuristically.

The potential energy (\ref{poten}) was briefly analized and the equilibria
points and the stable/unstable cases were obtained by means of the
usual derivative criteria, in accord with the other method (see \cite{murray}).
Also, using the virial theorem it was possible to write an expression
for the total mean energy of the system, equation (\ref{etotal}),
and a relation between the mean-squared velocity and the mean-linear
velocity, equation (\ref{velcmed}). 

It is not surprising to note that the Malthusian limit is trivially
obtained for the particular case when $B$ tends to $\infty$. Thus,
the potential energy for this case can be written as\begin{equation}
V_{Malthus}=-\frac{1}{2}A^{2}q^{2}\end{equation}

We have in this work studied the thus-called second order approach.
There is, certainly, a first order method of the Hojman et al. proceedure
\cite{Hojman-Urrut,Hojman}; we show briefly in the Appendix the way
an s-equivalent Lagrangean for the VLE can be constructed using this
approach. Even though the Lagrangean reproduces the logistic equation,
as required, the physical interpretation of its terms is not so clear.
However, generally speaking, it is important to bear in mind this
proceedure because it sometimes provides a variational formalism of
the physical phenomenon while the second order approach does not \cite{Hojman-Urrut}.

Finally, it could be useful to study in future works generalized auto-regulated
models as for example those reported in references \cite{Foerster,Hoerner,Kapitza,Kapitza99,Kobelev}.
It is also interesting to extend this work to the study of the quantum
systems related with self-regulated phenomena \cite{Munoz}.

\begin{acknowledgments}
The authors wish to thank A. Rincón, A. Skirzewski, J.L. Flores, J.
Toro-Mendoza, R.O. Rodríguez and D. Montiel for helpful comments.
ÁGMS aknowlegde a Centro de Estudios de Postgrado (CEP) of the Universidad
de Los Andes postgraduate student research grant. This work was partially
supported by División de Investigación, Facultad de Ciencias, La Universidad
del Zulia.
\end{acknowledgments}
\appendix
\begin{center}\textbf{APPENDIX}\end{center}

In order to obtain an s-equivalent Lagrangean for the Verhulst's equation
using the first order approach of the Hojman et al. method we proceed
as follows.

First, let write the logistic equation as\begin{equation}
\overset{.}{q}=kq\left(B-q\right)\label{a1}\end{equation}
 where $q\equiv q(t)$ is the generalized coordinate. Then, let write
and choose a new integral variable $Q(t)$\begin{equation}
Q(t)\equiv\int q(t)dt\end{equation}
 such that equation (\ref{a1}) transforms in\begin{equation}
\overset{..}{Q}(t)=k\overset{.}{Q}(t)\left[B-\overset{.}{Q}(t)\right]\end{equation}

This second order equation can be casted in two first order equations

\begin{equation}
\left\{ \begin{array}{c}
\overset{.}{x_{1}}=x_{2}\\
\overset{.}{x_{2}}=kx_{2}(B-x_{2})\end{array}\right.\end{equation}
 where $x_{1}\equiv x_{1}(t)$ and $x_{2}\equiv x_{2}(t)$. Note that
indeed $x_{2}$ is exactly the generalized coordinate $q$.

The solution of this system of equations is given by\begin{equation}
x_{1}=\int x_{2}dt+C_{1}\label{x1}\end{equation}

\begin{equation}
x_{2}=\frac{B}{1+BC_{2}e^{-kBt}}\label{x2}\end{equation}
 and this functional form for $x_{2}$ is just another way to write
the usual form (\ref{ver}). Now, solving (\ref{x1}) and (\ref{x2})
for $C_{1}$ and $C_{2}$ we obtain\begin{eqnarray}
C_{1} & = & kx_{1}-kBt-\ln B-\ln x_{2}\\
C_{2} & = & \frac{e^{kBt}}{x_{2}}\left(1-\frac{x_{2}}{B}\right)\end{eqnarray}
 and following Hojman and Urrutia \cite{Hojman-Urrut} the s-equivalent
Lagrangean is finally written as\begin{eqnarray}
L & = & e^{kBt}\left[\frac{k\overset{.}x_{1}}{x_{2}}-\frac{k\overset{.}x_{1}}{B}-\frac{kB}{x_{2}}+\frac{\overset{.}x_{2}}{x_{2}^{2}}-\frac{\overset{.}x_{2}}{Bx_{2}}+\right.\nonumber \\
 &  & -\frac{kB\ln(x_{2})}{x_{2}}+k^{2}x_{1}+k\ln n(x_{2})+\frac{k\overset{.}x_{2}x_{1}}{x_{2}^{2}}+\nonumber \\
 &  & -\frac{k^{2}Bx_{1}}{x_{2}}-\frac{k^{2}B^{2}t}{x_{2}}+\frac{kB\ln(B)}{x_{2}}+\nonumber \\
 &  & \left.-\frac{kBt\overset{.}x_{2}}{x_{2}^{2}}-\frac{\ln(B)\overset{.}x_{2}}{x_{2}^{2}}+\frac{\ln(x_{2})\overset{.}x_{2}}{x_{2}^{2}}\right]\end{eqnarray}

\newpage
Figure captions.

1. The Verhulst logistic equation for fixed carrying capacity ($B=20$).

2. The Verhulst potential for differents $k$ ($B=5$).

\begin{thebibliography}{10}
\bibitem{Van}V\'{a}n, P. \& Ny\'{\i}ri, B., \textit{Ann. Phys. (Leipzig)}, \textbf{8},
4, 331-354, 1999. 
\bibitem{Veinberg}Vainberg, M.M., Methods for the Study of Nonlinear Operators, (Holden-Day,
San Francisco, California, 1964). 
\bibitem{Tonti}Tonti, E., \textit{Int. Jour. Engin. Sci.,} \textbf{22}, 1343, 1984. 
\bibitem{Ichi}Ichiyanagi, M., \textit{Phys. Rep.,}\textbf{243/3}, 125, 1994. 
\bibitem{VanMu}V\'{a}n, P. \& Muschik, W., \textit{Phys. Rev. E}, \textbf{5/4},
3584, 1995. 
\bibitem{Nyi}Ny\'{\i}ri, B., \textit{Journal of Non-Equilibrium Thermodynamics,}
\textbf{16}, 217, 1991. 
\bibitem{Hojman-Urrut}Hojman, S., Urrutia, L.F., \emph{J. Math. Phys.,} \textbf{22}, 1896,
1981. 
\bibitem{HojmaShep}Hojman, S., Shepley, L.C., \textit{Rev.Mex.F\'{\i}s.}, \textbf{28},
149, 1982. 
\bibitem{Helm}Helmholtz, H., \textit{Journal f\"{u}r die reine und angewandte Mathematik}
(Berlin), \textbf{100}, 137, 1887. 
\bibitem{Hojman}Hojman, R., et al., \textit{Phys. Rev. D.,}\textbf{28}, 6, 1333-1336,
1983. 
\bibitem{verhulst}Verhulst, P.F., \textit{Corr. Math.et Phys.}, \textbf{10}, 113-121,
1838. 
\bibitem{murray}Murray, J.D.\emph{,} Mathematical Biology, 2nd Edition, (Springer
Verlag, Berlin, 1993). 
\bibitem{malthus}Malthus, T.R., An essay on the Principal of Population, 1798, (Penguin
Books, 1970). 
\bibitem{darcy}Thompson, D.W., On Growth and Form, Revised Edition, Dover Pub., New
York, 1992. 
\bibitem{Aonso}Alonso, M. and Finn, E.J., Fundamental University Physics, Volume
III, Quantum and statistical Physics (Addison-Wesley Publishing Company,
Reading, Massachusetts, 1968), p. 344-345. 
\bibitem{Nelson}Nelson, D.R. et al., \emph{Physical Review E}, \textbf{58}, No. 2,
1383-1403, 1998. 
\bibitem{Euler}Euler, L., \textit{Acta Acad. Sci. Imp. Petropol.}, 163-193, 1778. 
\bibitem{Town}Townsend, J.S., Electricity in Gases (Oxford University Press, London,
1914). 
\bibitem{Joos}Joos, G., Theoretical Physics (Dover Publications, Mineola, New York,
1986), pp. 458-462. 
\bibitem{pardo}Pardo, F., \emph{J. Math. Phys.,} \textbf{30}, No. 9, 2054-2061, 1989. 
\bibitem{Landau}Landau, L. , Mecanica, (Ed. Mir, Rome, 1975), p. 42. 
\bibitem{Goldstein}Goldstein, H., Classical Mechanics\textit{,} 2nd Edition, (Adison-Wesley,
London, 1980), p. 411. 
\bibitem{Foerster}Foerster, Von H., et al., \textit{Science}, \textbf{132}, 1291, 1960. 
\bibitem{Hoerner}Hoerner, von S.J., \textit{British Interplanetary Society}, \textbf{28},
691, 1975. 
\bibitem{Kapitza}Kapitza, S.P., \textit{Uspehi Fisicheskjh Nauk (Russia)}, \textbf{166},
1, 63-79, 1966. 
\bibitem{Kapitza99}Kapitza, S.P., \textit{How Many People Lived, Live and are to Live
in the World. An essay on theory of growth of Humankind,} Moscow,
Inst. Phys. Problem RAS, 1999. 
\bibitem{Kobelev}Kobelev, L. Ya., et al., \textit{Will the Population of Humanity in
the Future be Stabilized?}, http://arXiv.org/abs/physics/0003035,
16 Mar 2000. 
\bibitem{Munoz}Mu\~{n}oz S., A.G., Sierra P., D., Soldovieri, T., Rodriguez, O.,
\emph{Quantization of Self-Regulated Systems} (in preparation), 2003.
\end{thebibliography}
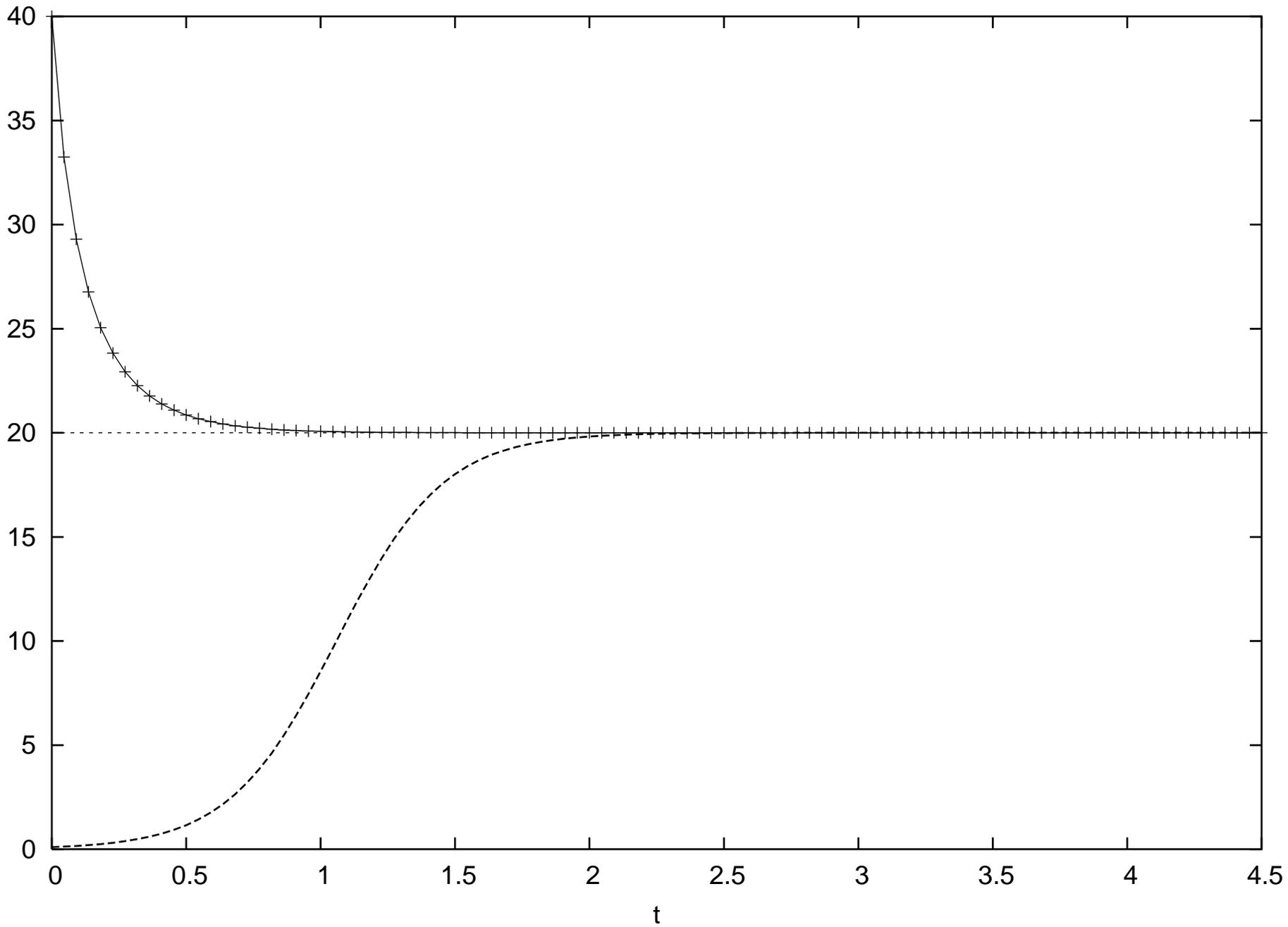
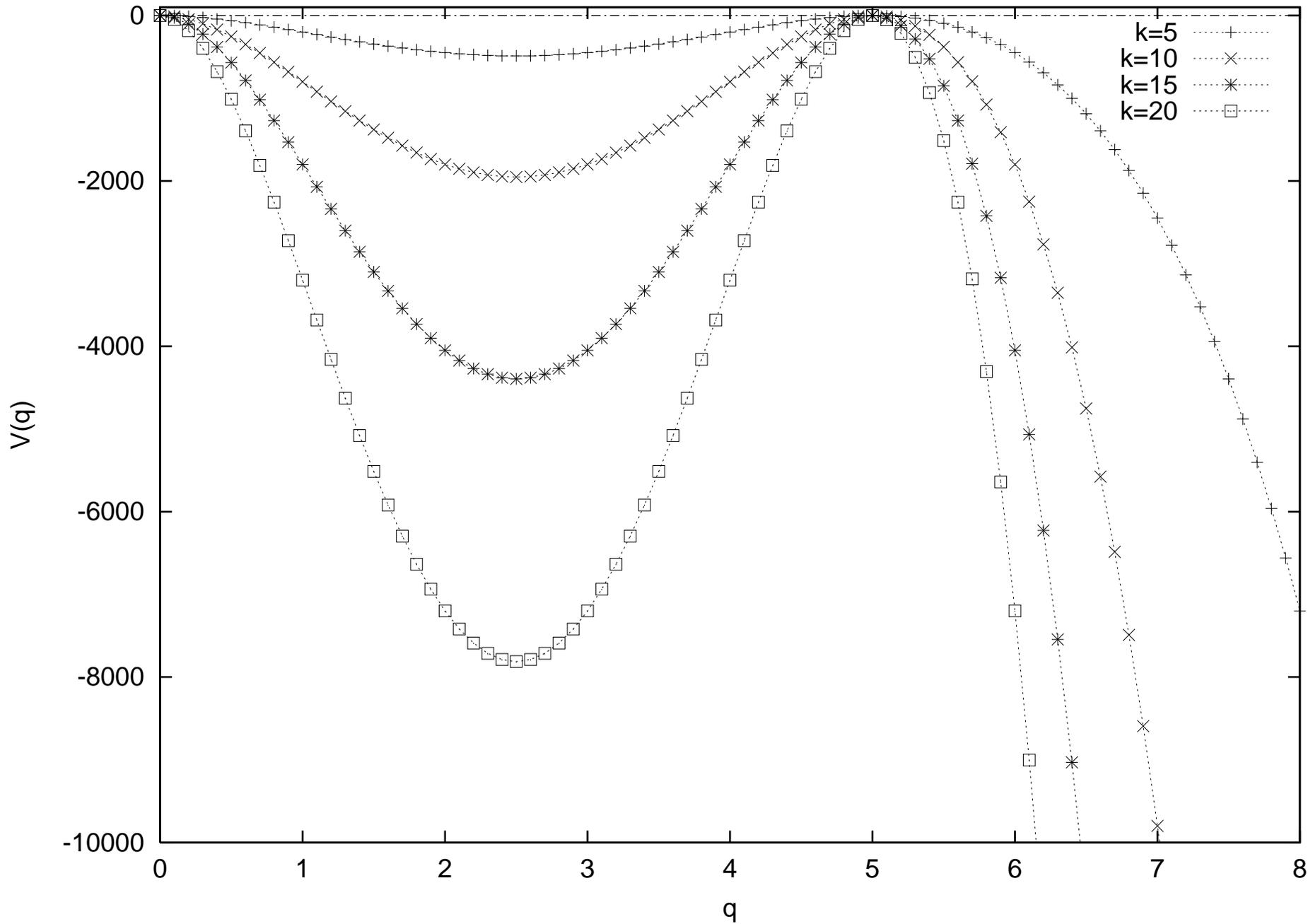
\end{document}